\documentclass[a4paper,11pt]{article}
\pdfoutput=0 % if your are submitting a pdflatex (i.e. if you have
\usepackage{jinstpub} % for details on the use of the package, please
\usepackage{color}% for personal comment

\title{\boldmath Performance Evaluation of the Analogue Front-End and ADC Prototypes for the Gotthard-II Development}

\author[a,1]{Jiaguo Zhang,\note{Corresponding author.}}
\author[a]{Marie Andr\"a,}
\author[a]{Rebecca Barten,}
\author[a]{Anna Bergamaschi,}
\author[a]{Martin Br\"uckner,}
\author[a]{Roberto Dinapoli,}
\author[a]{Erik Froejdh,}
\author[a]{Dominic Greiffenberg,}
\author[a]{Carlos Lopez-Cuenca,}
\author[a]{Davide Mezza,}
\author[a]{Aldo Mozzanica,}
\author[b]{Marco Ramilli,}
\author[a]{Sophie Redford,}
\author[a]{Marie Ruat,}
\author[a]{Christian Ruder,}
\author[a]{Bernd Schmitt,}
\author[a]{Xintian Shi,}
\author[a]{Dhanya Thattil,}
\author[a]{Gemma Tinti,}
\author[b]{Monica Turcato,}
\author[a]{Seraphin Vetter}

\affiliation[a]{Paul Scherrer Institut,\\5232 Villigen, Switzerland}
\affiliation[b]{European X-ray Free-Electron Laser Facility GmbH,\\Holzkoppel 4, 22869 Schenefeld, Germany}

\emailAdd{jiaguo.zhang@psi.ch}

\abstract{Gotthard-II is a silicon microstrip detector developed for the European X-ray Free-Electron Laser (XFEL.EU). Its potential scientific applications include X-ray absorption/emission spectroscopy, hard X-ray high resolution single-shot spectrometry (HiREX), energy dispersive experiments at \mbox{4.5 MHz} frame rate, beam diagnostics, as well as veto signal generation for pixel detectors. Gotthard-II uses a silicon microstrip sensor with a pitch of 50 $\mu$m or 25 $\mu$m and with 1280 or 2560 channels wire-bonded to readout chips (ROCs). In the ROC,  an adaptive gain switching pre-amplifier (PRE), a fully differential Correlated-Double-Sampling (CDS) stage, an Analog-to-Digital Converter (ADC) as well as a Static Random-Access Memory (SRAM) capable of storing all the 2700 images in an XFEL.EU bunch train will be implemented. Several prototypes with different designs of the analogue front-end (PRE and CDS) and ADC test structures have been fabricated in UMC-110 nm CMOS technology and their performance has been evaluated. In this paper, the performance of the analogue front-end and ADC will be summarized.
}


\begin{document}
\maketitle
\flushbottom

%\linenumbers

\section{Introduction}
\label{sec:intro}

The European X-ray Free-Electron Laser (XFEL.EU) \cite{Altarelli2006} in Hamburg-Schenefeld has been in operation and available for users since September 2017. First user experiments, for example serial femtosecond crystallography and study of single excited state of copper complex using pump-probe diffuse X-ray scattering and pump-probe valence-to-core X-ray emission spectroscopy, have been performed using AGIPD~\cite{Beat2011, Xintian2012}, LPD~\cite{Hart2012} and Gotthard-I~\cite{Aldo2011} detector\footnote{The XFEL.EU currently provides 30 X-ray pulses at 1.125 MHz in a train for first user experiments. The Gotthard-I runs at $\leq$ 800 kHz in burst mode with an image depth of 128. It is possible to measure X-ray photons from every two pulses for the current bunch structure but not enough in the future for the target bunch structure of 2700 pulses of \mbox{4.5 MHz} in a train.}. The XFEL.EU delivers ultrashort (< 100 fs), high intensity ($10^{12}$ ph/pulse) and fully coherent X-ray pulses with a peak brilliance $\sim$ 8 orders of magnitude higher than any other synchrotron radiation source. It is impossible to use photon-counting detectors at the XFEL.EU instead charge-integrating ones with a large dynamic range must be used. The unique bunch structure of the XFEL.EU, with a spacing of 220 ns in-between the 2700 pulses in a train and a train repetition rate of 10 Hz, poses great challenges to detectors in terms of a high frame rate of 4.5 MHz and requires a local storage of the images in the readout chip (ROC).

Gotthard-II is a charge-integrating microstrip detector for the XFEL.EU. To meet the strict requirements at the XFEL.EU, the development of the Gotthard-II detector is focused on: (a) a high speed, low noise, dynamic gain switching pre-amplifier (PRE) to cope with the 4.5 MHz frame rate and a large dynamic range up to $10^{4} \times 12.4$ keV photons in the meanwhile maintaining a single photon resolution with a Signal-to-Noise Ratio (SNR) > 10; (b) an on-chip Analog-to-Digital Converter (ADC) and a Static  Random-Access Memory (SRAM) capable of compact storage of all images due to the 2700 pulses in a bunch train; (c) an on-chip digital comparison circuit to generate veto signals for pixel detectors, for example AGIPD and LPD, which are only able to record 352 and 512 images per bunch train, respectively. With the veto signals, the memory of the pixel detectors storing "bad" images due to poor interactions between X-ray pulse and sample or due to unqualified X-ray pulses can be re-used for the other forthcoming pulses in the same train. The detailed specifications of the Gotthard-II detector and development strategy can be found in \cite{Zhang2017, Monica2014}.

\begin{figure}
\small
\centering
\caption{The architecture of the Gotthard-II ROC.}
\includegraphics[width=150mm]{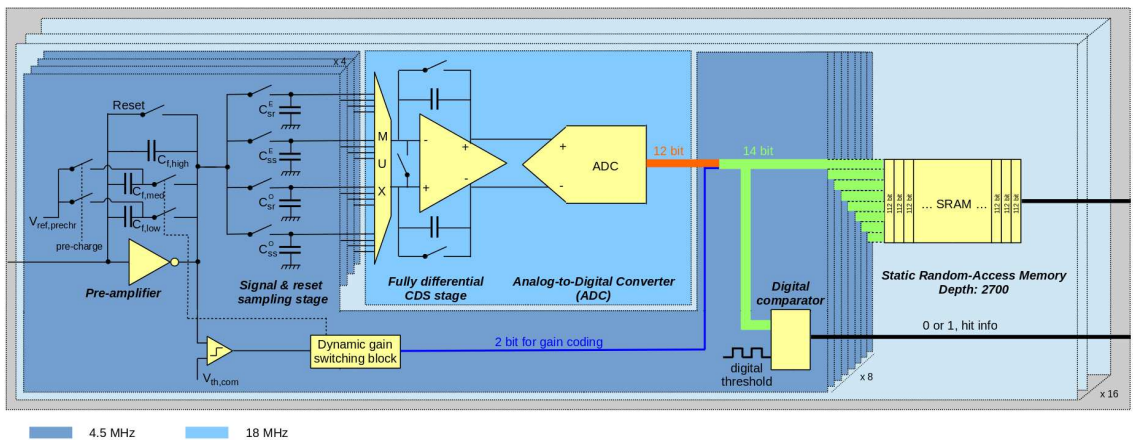}
\label{Architecture}
\end{figure}

The Gotthard-II detector uses a silicon microstrip sensor with a pitch of 50 $\mu$m or 25 $\mu$m and with 1280 or 2560 channels wire-bonded to the readout chip (ROC). As seen in figure~\ref{Architecture}, the Gotthard-II ROC includes an adaptive gain switching pre-amplifier (PRE), a fully differential Correlated-Double-Sampling (CDS) stage, a 12-bit Analog-to-Digital Converter (ADC), a Static Random-Access Memory (SRAM) with a depth for 2700 images, as well as a digital comparison circuit. The PRE output of each channel is connected to a "Signal \& reset sampling stage" which consists of two sets of analogue storage cells. In each set of analogue storage cells, one storage cell is used to record the output of the PRE immediately after reset while the other stores the additional signal induced by the incoming X-ray photons. The outputs of the "Signal \& reset sampling stage" of every four channels are multiplexed to one fully differential CDS stage and one ADC. Signals on the two analogue storage cells are subtracted and amplified by the differential CDS stage so that the CDS differential output is proportional to the integrated charge from the X-ray photons. Two sets of analogue storage cells are required in each channel because the signals in the Gotthard-II ROC follow a pipeline processing: When the signal generated by bunch-$i$ is being integrated by the PRE and stored into one set of analogue storage cells, the signal generated by bunch-($i-1$) stored in the other set of analogue storage cells is being "processed" by the CDS and ADC, while at the same time the output of the ADC after converting the signal generated by bunch-$(i-2)$ is being stored into the SRAM. The PRE and digital comparator of each channel runs at 4.5 MHz, while the CDS and ADC at \mbox{18 MHz} sampling rate, since one ADC is shared between 4 channels. The images saved in SRAM are readout in-between the bunch train spacing of the XFEL.EU of 99.4 ms. The output of the digital comparator as the veto signal is readout every \mbox{220 ns} from the ROC.

For Gotthard-II development, several ROC prototypes have been fabricated and tested. This paper will discuss the performance of the analogue front-end, including PRE and CDS, as well as the ADC prototypes in separated chapters. Further improvements are suggested based on the investigation as well.

\section{The analogue front-end prototypes}

A few analogue front-end prototypes, Gotthard-1.4, 1.5 \& 1.7, have been designed and fabricated in UMC-110 nm CMOS technology. Gotthard-1.4 \& 1.5 prototypes are designed based on the analogue front-end of Jungfrau~\cite{Aldo2014} and Gotthard-I~\cite{Aldo2011}, featuring a pre-chargeable dynamic gain switching PRE with three gain stages (noted G0, G1 and G2 with increasing feedback capacitance) and a single-ended CDS stage which can be by-passed after gain switching. The working principle of such PRE and CDS has been described elsewhere~\cite{Aldo2011, Aldo2014, Zhang2017}. The storage of signals in Gotthard-1.4 \& 1.5 is done in analogue memory cells during charge integration and the stored signals of all channels are readout through strip buffers and an off-chip driver. The performance of the Gotthard-1.4 \& 1.5 prototypes has been investigated in detail and reported in~\cite{Zhang2017}. It was found that the coupling effect in-between strip channels is not negligible: The coupling factor, defined as the charge measured by the neighbouring strip channel divided by the charge measured by the strip channel collecting all charge carriers in the silicon sensor produced by X-ray photons, is about $\sim$ 6.2\%; in addition, the cross-talk at the gain switching point, i.e. the signal induced to the neighbouring strip channel when the central one switches, is \mbox{$\sim$ 3.3$\times$12.4 keV} photons for a single neighbouring channel and \mbox{$\sim$ 9.8$\times$12.4 keV} photons for up to three neigbouring channels. Both effects increase the complexity of detector calibration. 

To overcome the coupling-related problems observed in Gotthard-1.4 \& 1.5, a high DC gain PRE has been designed and implemented in the Gotthard-1.7 prototype. Instead of using a cascaded push-pull inverter in the PRE, a split-transistor push-pull inverter with a high DC gain and low power consumption was developed. The DC gain is found to be 600-900 from the simulation for a working voltage range from 1.2 V to 1.4 V, and the power consumption is < 0.25 mW. In addition, the fully differential CDS stage and analogue memory cells in between PRE and CDS, planned for Gotthard-II as seen in figure~\ref{Architecture}, have been designed and implemented into Gotthard-1.7 as well, for the convenience of future integration together with the ADC and the SRAM. In the Gotthard-1.7 prototype, the analogue signals from the output of the fully differential CDS stage are still read out with a multiplexed serial output with a fully differential off-chip driver, in a similar way of \mbox{Gotthard-1.4 \& 1.5}. The performance of Gotthard-1.7 prototype in terms of coupling effect, single-photon sensitivity, dynamic range and noise has been investigated and will be presented.

\subsection{Coupling effect}

\begin{figure}
\small
\centering
\caption{Capacitive coupling effect in the Gotthard-1.7 prototype: (a) Correlation map of energy measured by two neighbouring strip channels using low-rate X-ray fluorescence; (b) Cross-talk at the gain switching point determined from a measurement using infrared laser.}
\includegraphics[width=73.5mm]{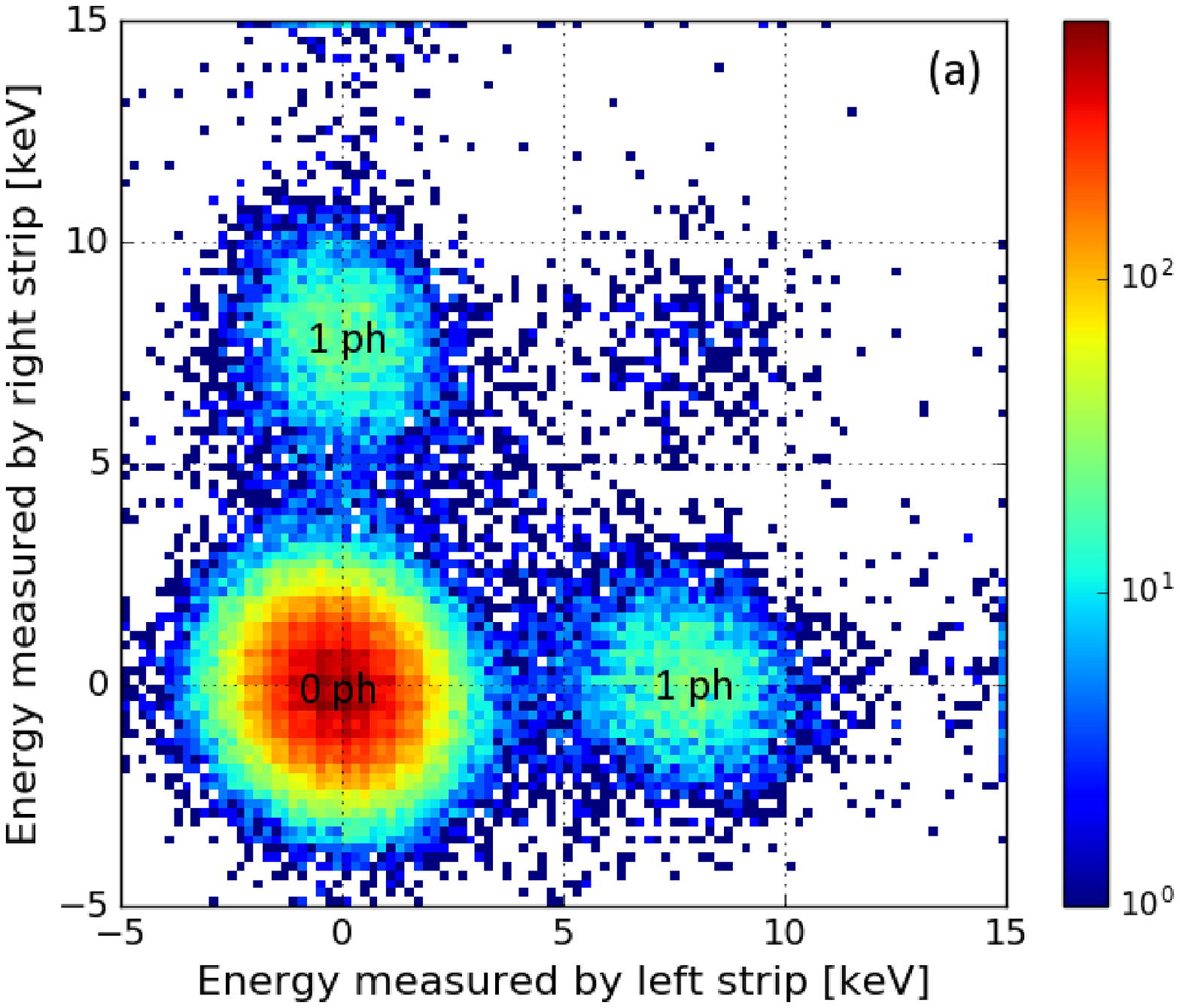}
\includegraphics[width=75mm]{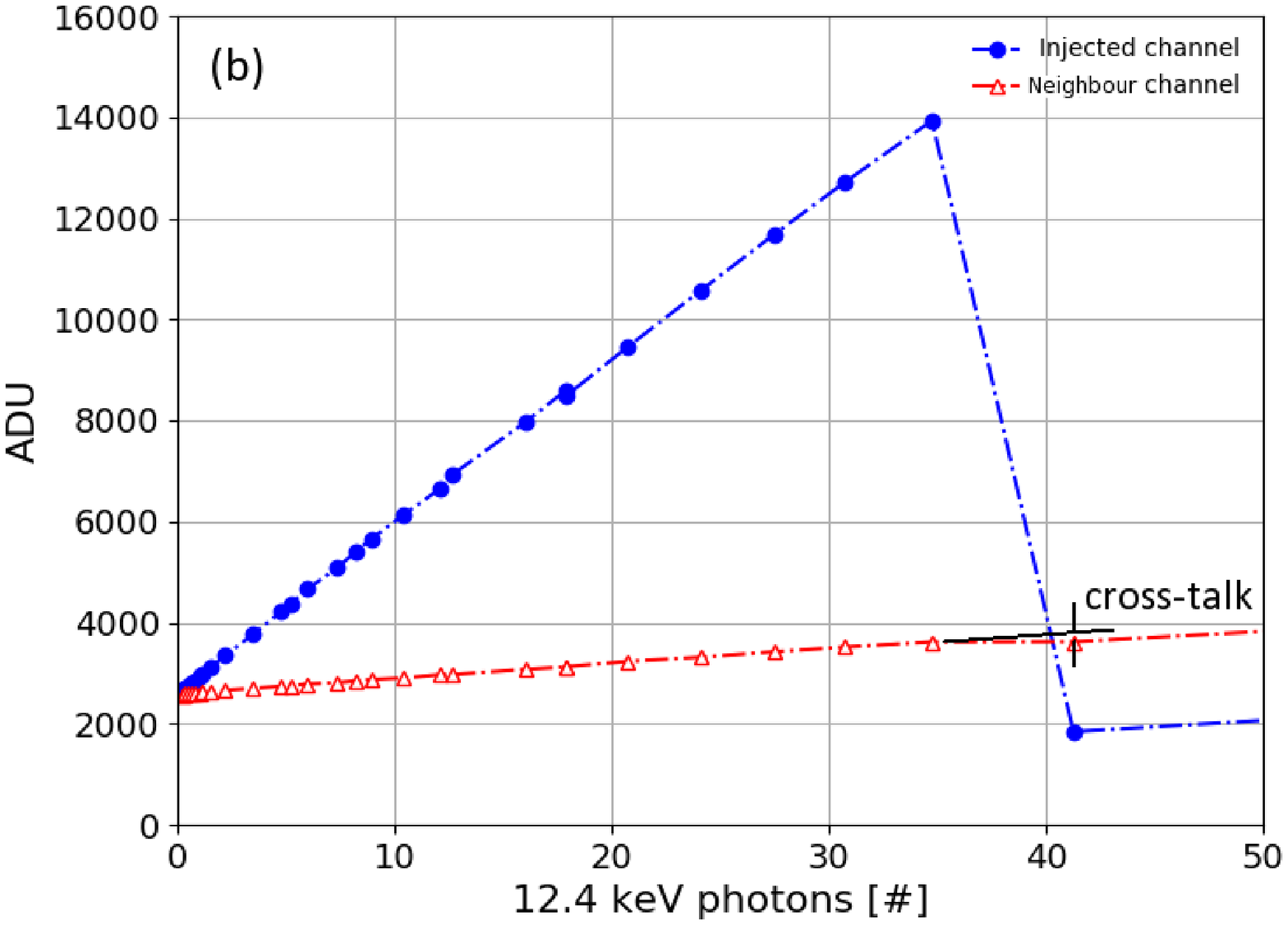}
\label{Coupling}
\end{figure}

The capacitive coupling before gain switching (in G0: the highest gain with lowest capacitance used in the feedback loop of the PRE) between neighbouring strip channels has been investigated through a measurement using X-ray fluorescence with a low rate ($\leq$ 1 photon per frame per channel). The X-ray fluorescence of 8.05 keV was generated by a commercial X-ray tube using a secondary copper target and was illuminated on the strip side of the sensor to minimize charge-sharing effect. The measured raw ADU values of all channels have been converted to energy after a pedestal correction and a gain correction. Figure~\ref{Coupling}(a) shows the correlation map of energy measured by two neighbouring strip channels. The three visible regions in the figure refer to 0-photon and 1-photon regions with single photon hit onto one of the two strip channels. The intermediate region linking the two 1-photon regions is due to charge sharing between two strip channels. By projecting the 1-photon region to $x$-axis and $y$-axis, the two distributions were fitted by a Gaussian function and the mean values from the Gaussian fit give the energies measured by the two strip channels when the charge generated by the single photon is fully collected by one of the two strip channels (no charge sharing and influence due to electronic noise excluded). The ratio of the energies measured by the two strip channels (channel o/w charge generated by the photon), namely the two mean values from the Gaussian fit, is the coupling factor. The coupling factor in this case is found to be $\sim$ 0.8\% thanks to the high DC gain of the PRE. Such a coupling is $\sim$ 1/8 of the factor measured from Gotthard-1.4 \& 1.5.

The cross-talk at the gain switching point is also determined: Infrared laser with different intensities was injected into the center of one strip until its PRE is switched to G1 from G0, while the change of the output of its neighbouring strip channel is measured. Figure~\ref{Coupling}(b) shows the output of the injected strip channel (in blue dots) and its neighbouring strip channel (in red open triangles) as function of laser intensities, which have been converted to the number of 12.4 keV photons. The gain switching of the injected strip channel happens at $\sim$ 35 photons. At the gain switching point, the output of the neighbouring strip channel is reduced by $\sim$ 150 ADU: The reduction can be explained by the redistribution of charge to the feedback capacitors of the PRE of the injected strip and to its neighbouring channels with reduced coupling effect due to the increase of equivalent input capacitance at the input node of the injected channel, which results in a reduction of charge division into neighbouring channels. The cross-talk at the gain switching point corresponds to \mbox{$\leq$ 0.5$\times$12.4 keV photons}, which is $\leq$ 1/7 of the cross-talk measured from the Gotthard-1.4 \& 1.5 prototypes with a lower DC gain. The improvement factor is consistent with the coupling value determined with single photon illumination.

\subsection{Single-photon sensitivity and dynamic range}
\label{subsect_ph}

\begin{figure}
\small
\centering
\caption{The single-photon sensitivity and dynamic range of the Gotthard-1.7 prototype.}
\includegraphics[width=75mm]{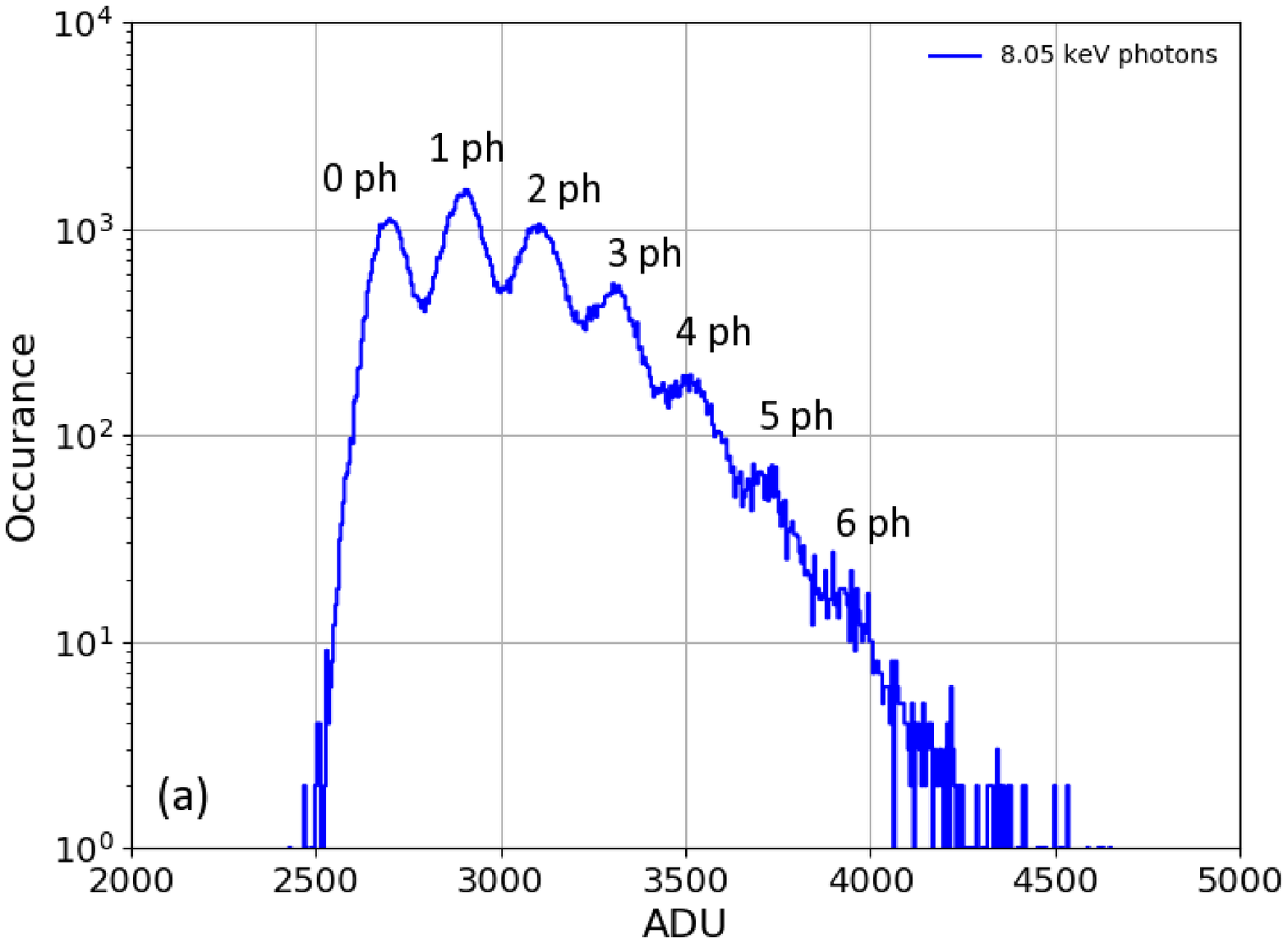}
\includegraphics[width=75mm]{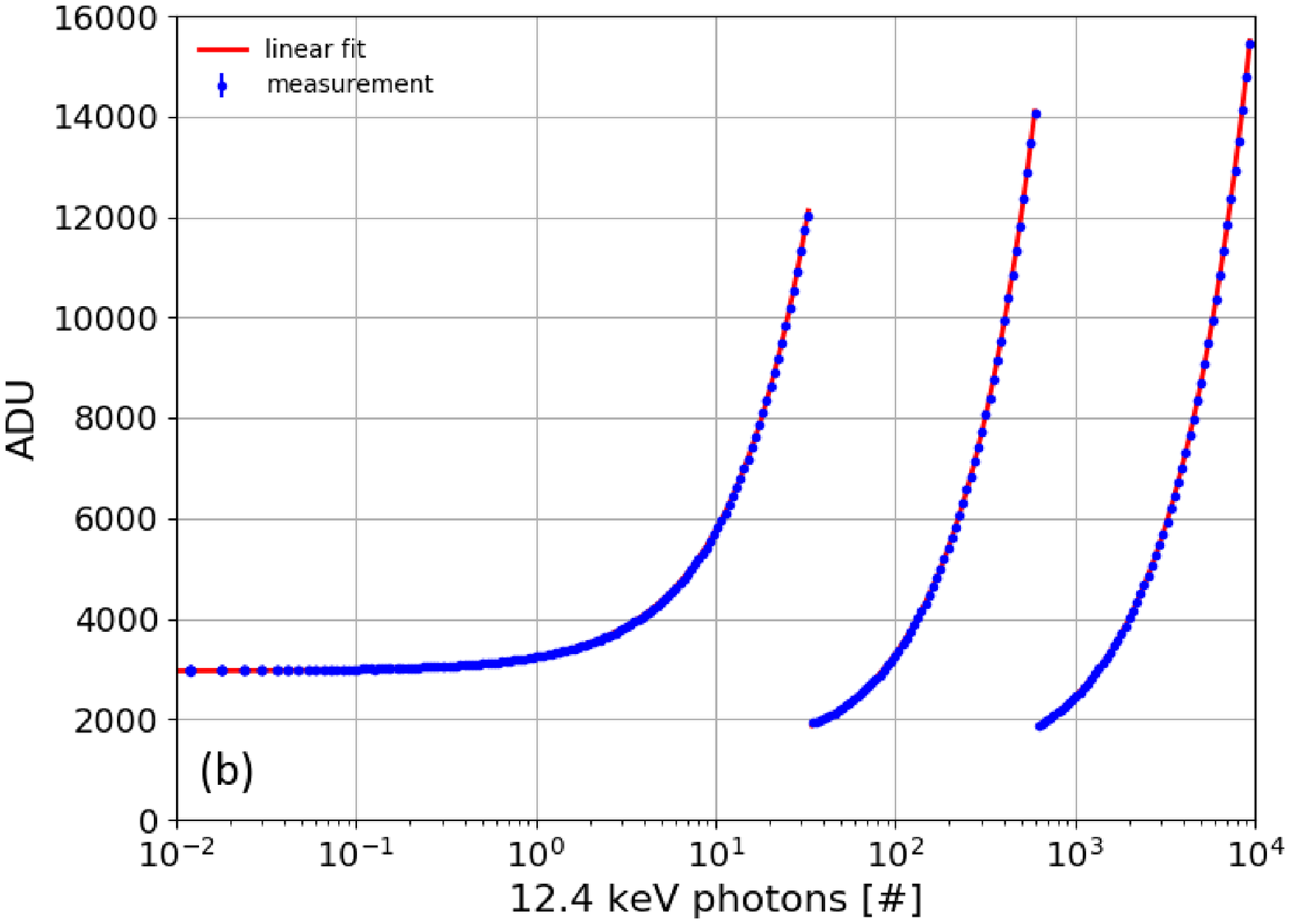}
\label{DR}
\end{figure}

Figure~\ref{DR}(a) shows the histogram of measured ADU values of one strip channel using the afore-mentioned X-ray set-up and fluorescence target but with a higher flux. Peaks for 0, 1,..., up to 6 coincident photon entries can be clearly seen and they are very well separated. Note that the $y$-axis is shown in a logarithmic scale. From the measurement, single-photon resolution for 8.05 keV has been obtained with a signal-to-noise ratio of > 5.

The dynamic range of the Gotthard-1.7 prototype is determined through a dynamic range scan using sensor backside pulsing. The method has been described in~\cite{Davide2016} for the calibration of the AGIPD detector, which is able to reach part of the range of G1 of the pixel detector. Due to the bigger sensor capacitance of each strip compared to the pixel, with the same voltage change/step caused by the external pulsing onto the silicon sensor, more charges are generated at the input node of the PRE of each strip channel and larger dynamic range is easily achieved with the method in strip detectors. Figure~\ref{DR}(b) shows the results of a dynamic range scan using sensor backside pulsing. The voltage step from the external pulse has been converted to number of 12.4 keV photons, which is shown as $x$-axis in the figure. The dynamic range has reached $10^{4} \times 12.4$ keV photons and gain switching happens at $\sim$ 35 and $\sim$ 650 photons to G1 and G2, respectively. Data of each gain stage has been fitted with a linear function (shown as the red line in the figure) so that linearity can be calculated. The maximum non-linearity in all gain stages is found to be less than 1\% of the signal.

\subsection{Noise}

\begin{figure}
\small
\centering
\caption{The conversion gain (G0) of the Gotthard-1.7 prototype: (a) Conversion gain over all channels; (b) Difference in conversion gain between "odd sampling" and "even sampling"; (c) Histogram of conversion gain showing gain variations.}
\includegraphics[width=70mm]{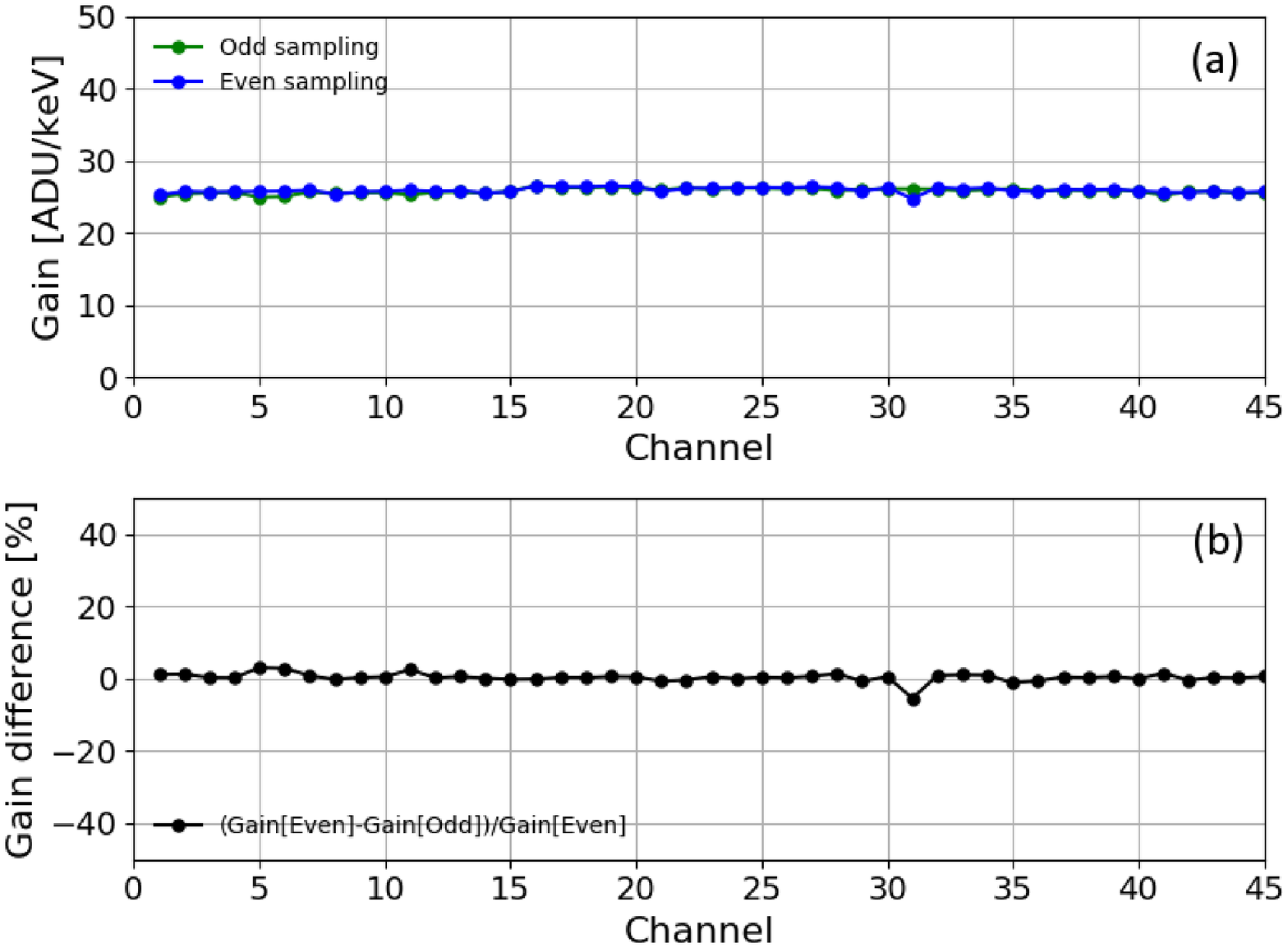}
\includegraphics[width=75mm]{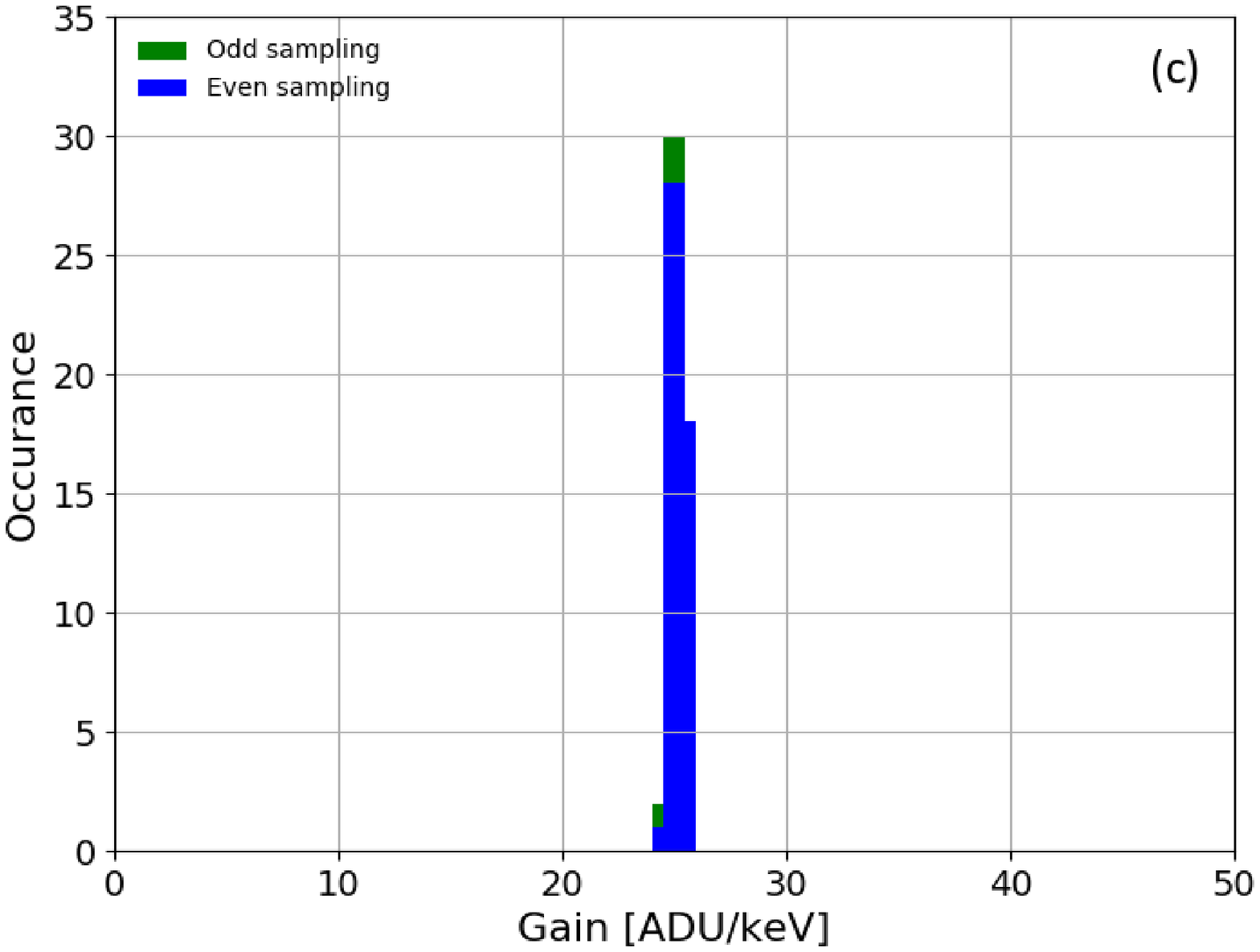}
\label{Gain}
\end{figure}

To evaluate the noise of the prototype, the conversion gain (G0) has to be determined in advance. The conversion gain (G0) was calculated based on the histogram shown in figure~\ref{DR}(a) for every strip channel. Taking the peak positions in terms of ADU values for each individual photon peak and plotting them as function of their corresponding photon energies, the slope in a linear fit gives the conversion gain in a unit of ADU/keV. Figure~\ref{Gain}(a) shows the conversion gain (G0) determined for all the investigated strip channels using one of the two sets of analogue storage cells (noted as "odd sampling" in green for one set, "even sampling" in blue for the other).  As shown in figure~\ref{Gain}(b), minor differences in the conversion gain (G0) between "odd sampling" and "even sampling" have been found, which indicates broadly consistent capacitance of the analogue memory cells between the two sets in the prototype production. Figure~\ref{Gain}(c) shows the histogram of the conversion gain (G0): the variation of the conversion gain between different channels is < 2\%.

\begin{figure}
\small
\centering
\caption{The noise of the analogue front-end prototypes: (a) Noise r.m.s. in electrons; (b) Difference in noise between "odd sampling" and "even sampling"; (c) Histogram of noise.}
\includegraphics[width=70mm]{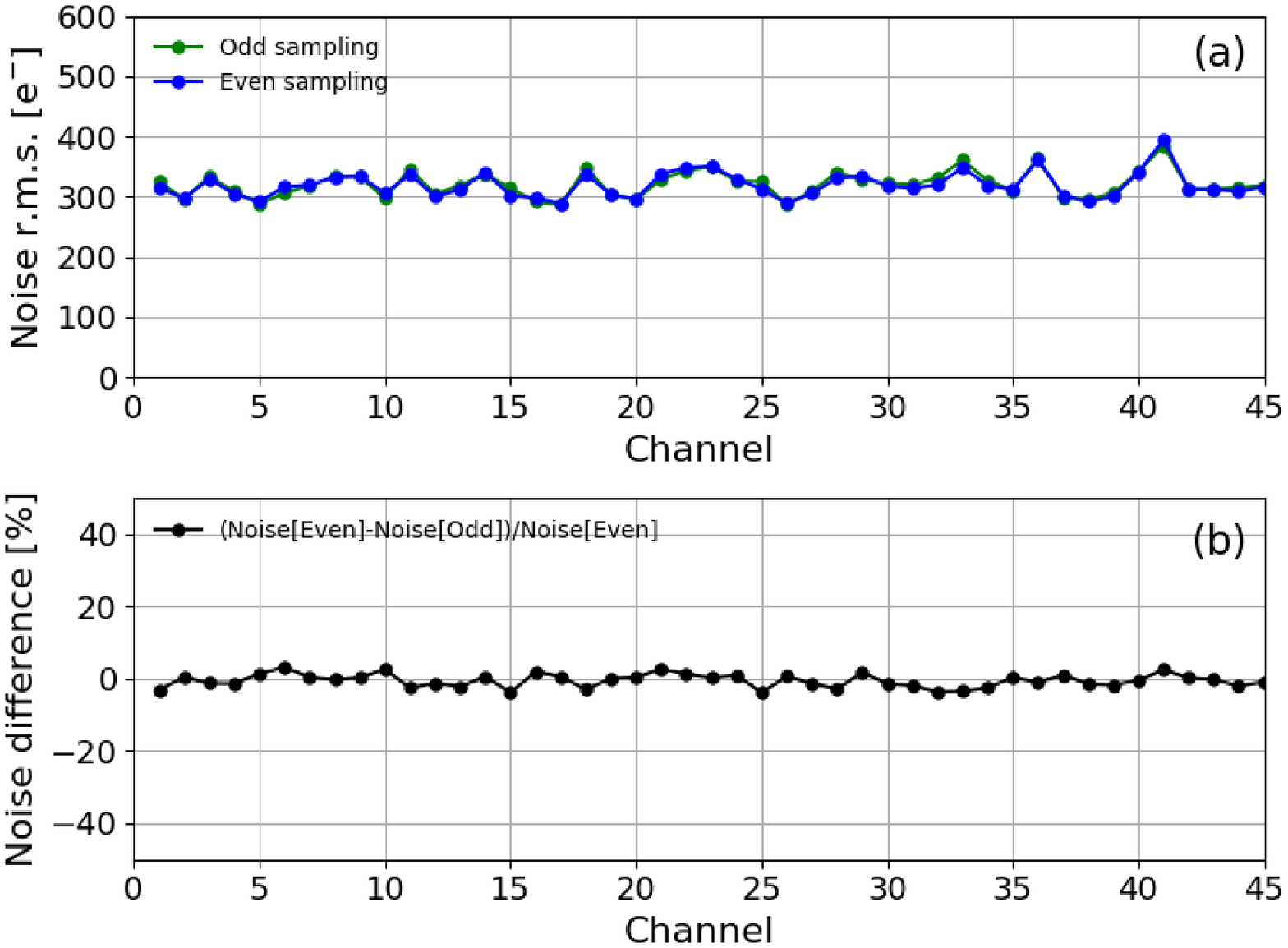}
\includegraphics[width=75mm]{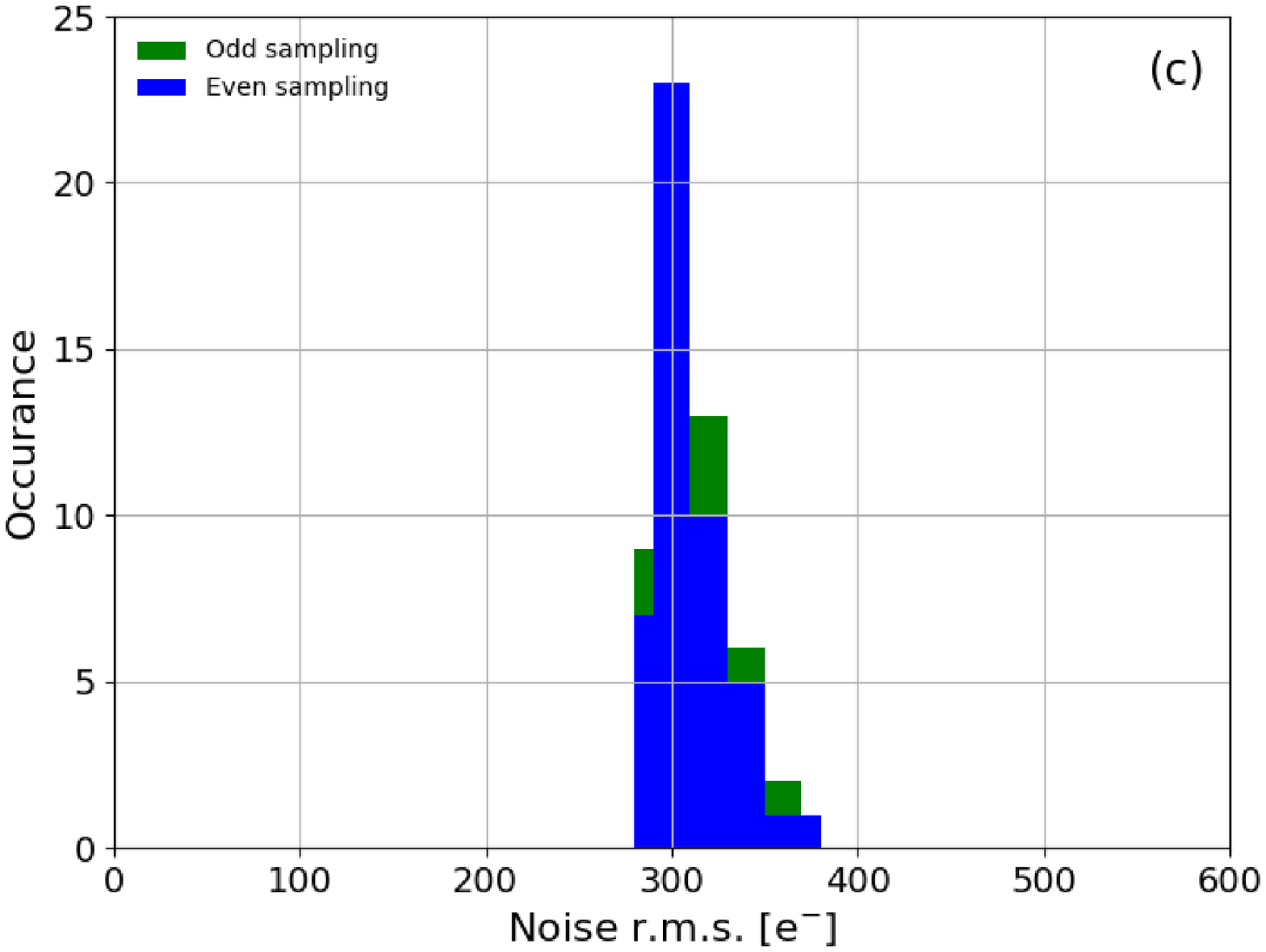}
\label{Noise}
\end{figure}

With the conversion gain, the equivalent noise charge has been calculated by: $\textrm{noise}\ r.m.s.\ [e^{-}] = \sigma\ [\textrm{ADU}]/gain\ [\textrm{ADU/keV}] \cdot 1000/3.6\ \textrm{eV}$, where $\sigma$ is the standard deviation of a Gaussian fit to the histogram of the measured ADU values obtained in a dark measurement, and 3.6 eV is the energy needed to generate one electron-hole pair in the silicon sensor. Figure~\ref{Noise}(a) and (b) are the determined noise r.m.s. for "odd sampling" and "even sampling" with an integration time of \mbox{5 $\mu$s}, and their difference in percentage. The nominal value of the noise is $\sim$ 300 e$^{-}$ with $\sim$ 10\% channel-to-channel variations as shown in figure~\ref{Noise}(c).

The noise in the Gotthard-1.7 prototype is acceptable for 12.4 keV X-ray photons with a good SNR of 10 and for 6 keV X-ray photons with a SNR of 5. However, in order to have a better single-photon resolution for even lower photon energies, new prototypes (Gotthard-1.8) with different transistor sizes of the PRE and different capacitance of the analogue memory cells have been designed and optimized, and their noise performance is going to be investigated.

\section{The ADC prototypes}

Gotthard-II requires a 10-bit Successive Approximation Register (SAR) ADC with a speed \mbox{> 18 MS/s}. The final ADC design will have 12-bit outputs: the 12-bit raw data is converted to 10-bit data using a look-up table obtained during calibration performed by injecting an input voltage ramp. The ADC accepts differential signals as inputs for the inherent benefits from noise and interference immunity. 

Two ADC prototype versions with different DAC array architectures have been designed and investigated. Both prototypes include the following main components: two sets of Digital-to-Analog Converter (DAC) arrays (when one is sampling the signal, the other is used for the conversion), one comparator, and one asynchronous control logic, which does not need any high frequency clock for each individual comparisons. When one DAC array is sampling the input signals, it is disconnected from the comparator, while the other DAC array is connected to the comparator. After each comparison, a "ready" signal is generated and sent to the control logic to control the switches in serial to the capacitors in the DAC array connected to the reference voltages, and thus changes the voltages at the input of the comparator for the next comparison. The 12-bit outputs are generated after 12 comparisons.

The first prototype, ADC-0.1, uses a conventional DAC array. The capacitance in the DAC array increases from Least-Significant-Bit (LSB) to Most-Significant-Bit (MSB) following a power law with a base of 2: $2^{i}C$ with $C$ the unit capacitance of 60 fF in the design. The total capacitance of the DAC array is $\sim$ 250 pF, which significantly reduces the speed of the ADC. For ADC-0.1, the maximum achievable speed is 10 MS/s, which is below the design target of 18 MS/s. To reduce the total capacitance of the DAC array and thus increase the speed, the ADC-0.2 prototype has been developed employing a split-capacitor DAC array instead \cite{Chang2013}; three versions of the DAC array have been produced with different unit capacitance of \mbox{20 fF}, 40 fF and 60 fF. The architecture and first test results of the ADC-0.2 prototype will be presented.

\subsection{Architecture of the DAC array of the ADC prototype}

\begin{figure}
\small
\centering
\caption{The split-capacitor DAC array of the ADC-0.2 prototype.}
\includegraphics[width=100mm]{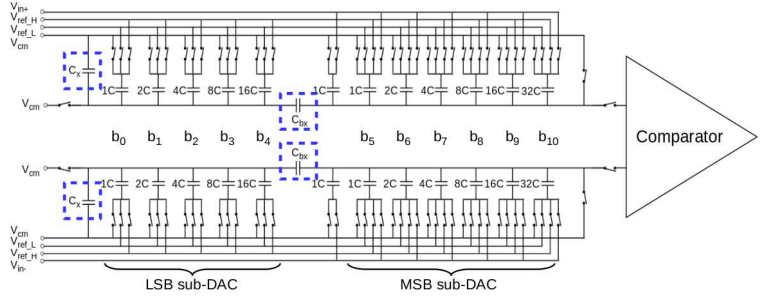}
\label{ADC_DAC}
\end{figure}

The architecture of the split-capacitor DAC array of the ADC-0.2 prototype is shown in figure~\ref{ADC_DAC}. It consists of two sub-DAC arrays at each input of the comparator: LSB sub-DAC and MSB sub-DAC, which are separated by a bridge capacitor, $C_{bx}$, whose capacitance is equal to a unit capacitance of $C$. In addition, a side capacitor labeled as $C_{x}$ is implemented in the LSB sub-DAC side. Due to the existence of the parasitic capacitance of $C_{bx}$ ($C_{bx} > 1C$ after fabrication), by tuning the capacitance of $C_{x}$, it is possible to match the equivalent capacitance of the LSB sub-DAC in serial to $C_{bx}$ with the minimum capacitance of the capacitor in the MSB sub-DAC; in this case, continuous output codes can be expected. The relation between $C_{x}$ and $C_{bx}$ can be found in \cite{Chang2013}. The capacitance of $C_{x}$ in our case varies according to the design with different unit capacitance and their values are given in table~\ref{Table_cap}. The capacitance values are indicated, in terms of number of unit capacitance, as well. In the signal sampling cycle, the DAC array is disconnected from the comparator and it is charged up by the input signals, $V_{in,+}$ and $V_{in,-}$, while keeping both sides of $C_{bx}$ connected to the common-mode voltage, $V_{cm}$. In the comparison cycle, the DAC array is disconnected from the input signals but connected to $V_{cm}$ in the meanwhile keeping the nodes at the two sides of $C_{bx}$ floating; thus, the input signals are written to the input nodes of the comparator, given by $2V_{cm} - V_{in,+}$ and $2V_{cm} - V_{in,-}$. After this step, the first comparison is made directly between the voltages at the two inputs of the comparator, $V_{comp,+}$ and $V_{comp,-}$, after which bit-11 ($b_{11}$) is generated. If the voltage at the "+" input of the comparator is smaller than its "-" input, $V_{comp,+}$ < $V_{comp,-}$, the capacitor for $b_{10}$ at the "+" input side is connected to the high reference voltage, $V_{ref\_H}$, while the capacitor at the "-" input to the low reference voltage, $V_{ref\_L}$: This increases $V_{comp,+}$ by $|V_{ref\_H}-V_{cm}|/2$ and reduces $V_{comp,-}$ by $|V_{cm}-V_{ref\_L}|/2$. Similarly, if $V_{comp,+}$ < $V_{comp,-}$, $V_{comp,+}$ is reduced by $|V_{cm}-V_{ref\_L}|/2$ and $V_{comp,+}$ increased by $|V_{ref\_H}-V_{cm}|/2$. Since the settings for $V_{ref\_H}$ and $V_{ref\_L}$ are usually symmetric and centered at $V_{cm}$, the changes of voltage at both inputs of the comparator are identical and given by $|V_{ref\_H}-V_{cm}|/2 = |V_{cm}-V_{ref\_H}|/2$. If there is no capacitance mismatch in the DAC array, after each comparison, the input signals of the comparator change by: $|V_{ref\_H}-V_{cm}|/2^{i}$, with $i$ the $i$-th comparison.

\begin{table}[htbp]
\centering
\begin{tabular}{|c|c|c|c|}

\hline
\textbf{Unit capacitance, $C$} & 20 fF & 40 fF & 60 fF \\
\hline
\textbf{Bridge capacitance, $C_{bx}$} & $1C$ & $1C$ & $1C$ \\
\hline
\textbf{Side capacitance, $C_{x}$} & $24.5C$ & $35.5C$ & $62.5C$ \\
\hline

\end{tabular}
\caption{The values of bridge capacitance and side capacitance used in designs for different unit capacitance.}
\label{Table_cap}
\end{table}

\begin{figure}
\small
\centering
\caption{Post-layout simulation for the ADC-0.2 prototype. From top to bottom: Input clock, signals at the input of the comparator, 12-bit outputs from the comparator, "ready" signals after each comparison.}
\includegraphics[width=125mm]{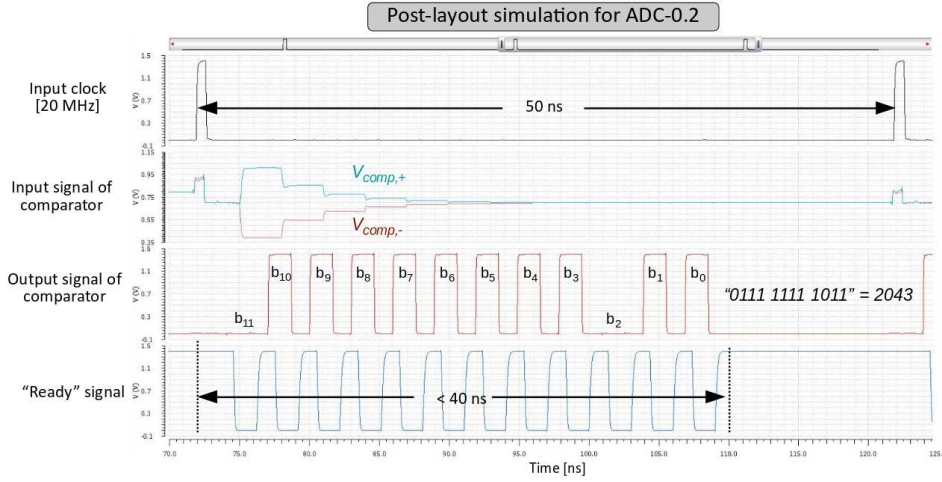}
\label{InputChange}
\end{figure}

As an example, figure~\ref{InputChange} shows results of a post-layout simulation for the ADC-0.2 prototype with an input clock of 20 MHz. In the simulation, a common-mode voltage of $V_{cm}$ = 0.7 V, \mbox{$V_{ref\_L}$ = 0 V}, $V_{ref\_H}$ = 1.4 V, input voltages of $V_{in,+}$ = 701.5 mV and $V_{in,-}$ = 698.5 mV are applied. It can be seen after each comparison, the changes of the input signal follows $|V_{ref\_H}-V_{cm}|/2^{i}$. It is found that the 12 "ready" signals with inversed polarity are sent out within 40 ns, which suggests a maximal speed of $\sim$ 25 MS/s.

\subsection{Performance of the ADC prototype}

\begin{figure}
\small
\centering
\caption{The noise performance of the ADC-0.2 prototype: (a) Noise in LSB; (b) Noise in mV.}
\includegraphics[width=100mm]{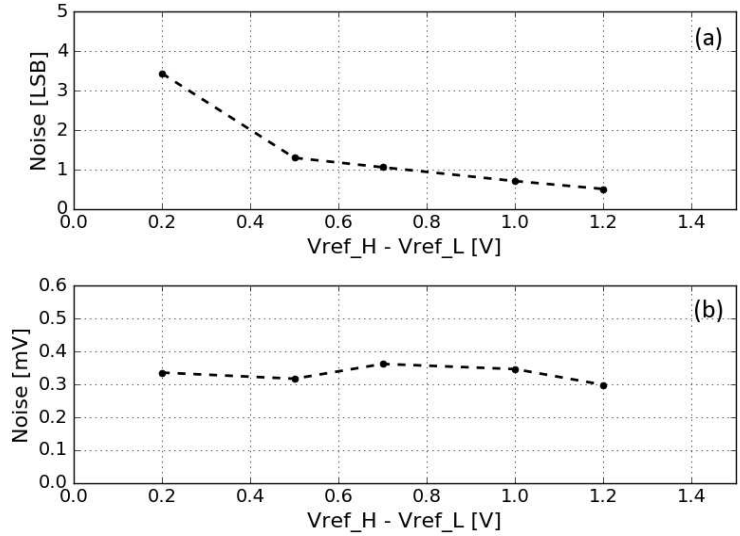}
\label{ADC_noise}
\end{figure}

The noise performance of the ADC-0.2 prototype has been evaluated for different reference voltages. In the measurement, $V_{cm}$ is set at 0.7 V and the reference voltage range, $|V_{ref\_H} - V_{ref\_{L}}|$, has been changed from 0.2 V to 1.2 V. The input signals of the ADC were generated by a low-noise on-chip differential buffer without any external input in order to distinguish the noise source from the external voltage supply. Figure~\ref{ADC_noise}(a) shows the measured noise in LSB (or ADU value) for different reference voltage ranges: It decreases with increasing range. In order to achieve a noise less than 1 LSB at 12-bit, the reference voltage $|V_{ref\_H} - V_{ref\_{L}}| \geq 1\ V$ is essential. The noise in LSB has been converted to a unit of mV using $\textrm{Noise[mV]} = |V_{ref\_H} - V_{ref\_{L}}|/2^{12} \cdot \textrm{Noise[LSB]}$, and shown in figure ~\ref{ADC_noise}(b). The intrinsic noise of the ADC-0.2 prototype is 0.3-0.4 mV.

\begin{figure}
\small
\centering
\caption{The response of the ADC-0.2 prototype to linear voltage input: (a) Direct output from the 12-bit data; (b) Output after calibration to 10-bit.}
\includegraphics[width=75mm]{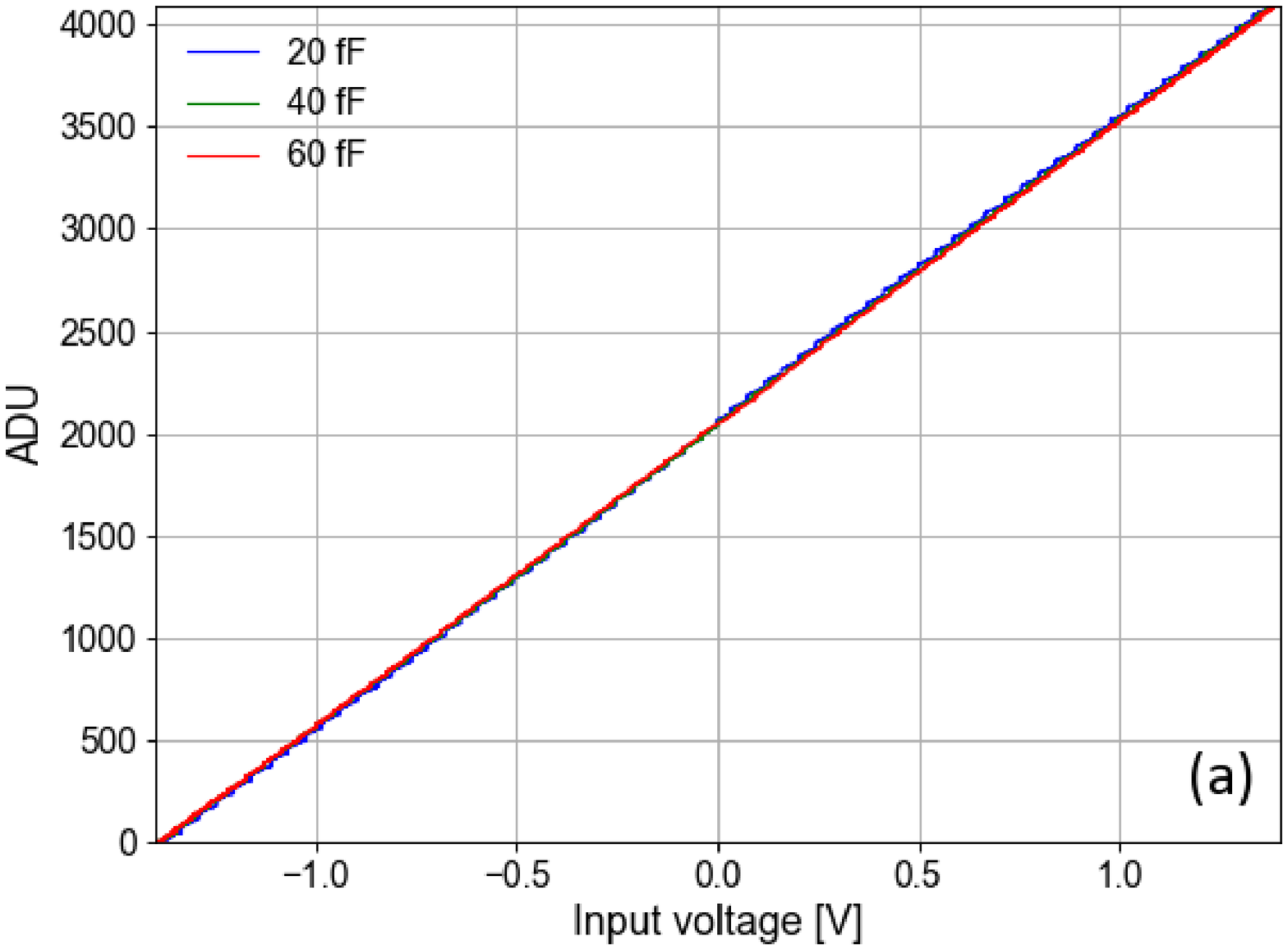}
\includegraphics[width=75mm]{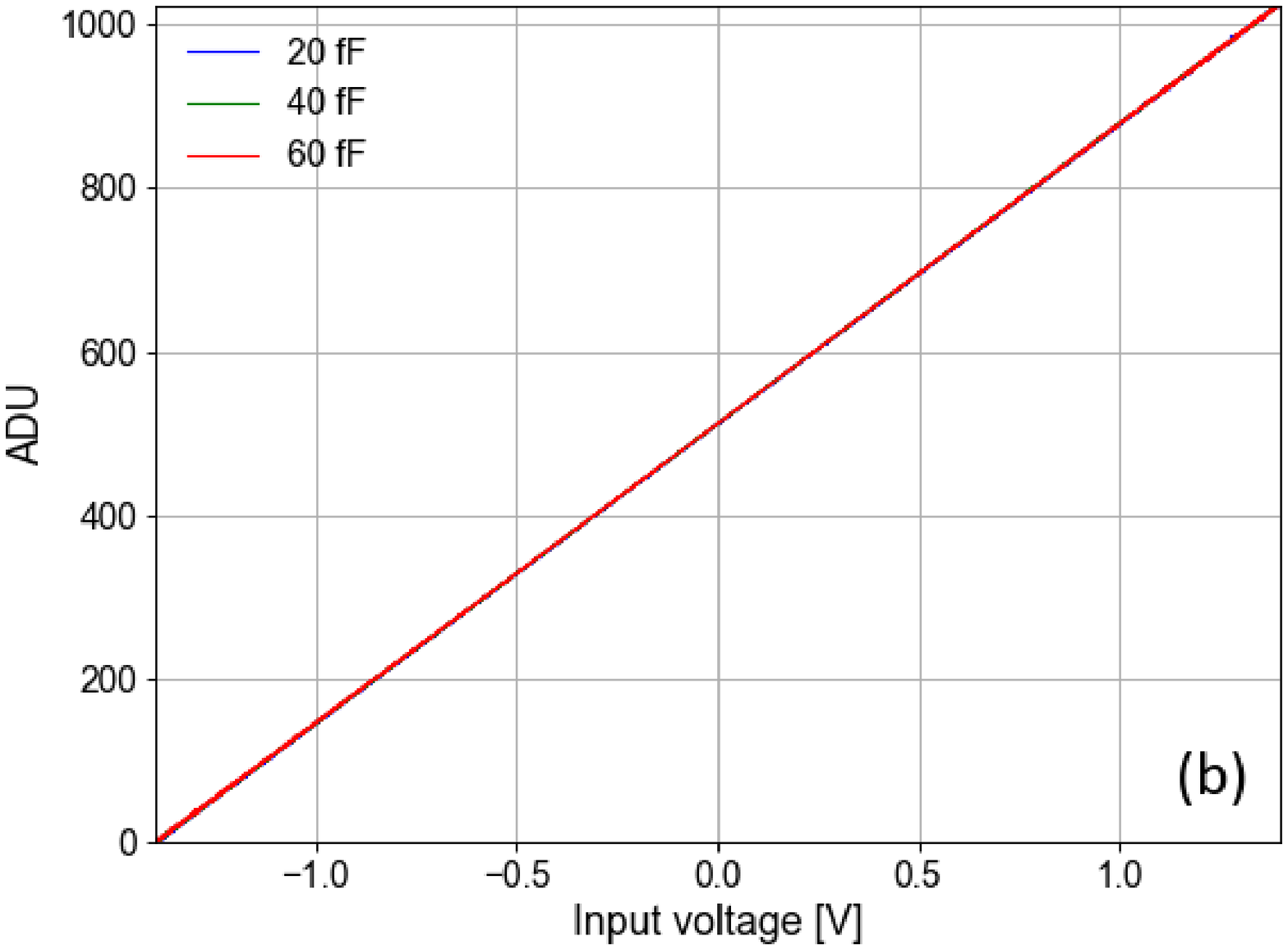}
\label{ADC_response}
\end{figure}

The linear response of the ADC-0.2 prototype with different unit capacitance of 20 fF, 40 fF and 60 fF have been tested using an external 18-bit DAC evaluation board, which provides relatively low-noise input signals for the ADC. Figure~\ref{ADC_response}(a) shows the ADC output as function of input voltage. It has been observed that certain number of missing codes appears periodically every 64 ADU codes for the three different unit capacitance. The number of missing codes per 64 ADU is 28 (43.75\%), 16 (25.0\%) and 12 (18.75\%) for 20 fF, 40 fF and 60 fF, respectively. It is attributed to the fact that the equivalent capacitance of $C_{bx}$ in serial to the LSB sub-DAC does not match the minimal capacitance in the MSB sub-DAC. The mismatch can be compensated by an adjustable capacitance of $C_{x}$, which has been implemented in the ADC-0.3 prototype and is currently under investigation.

To investigate the performance of the ADC-0.2 prototype at 10-bit, a simple calibration method has been applied: 1) The Differential Non-Linearity (DNL) at 12-bit is extracted from a histogram test through a slow ramp of the input voltage using the 18-bit DAC with a constant ramping speed; 2) The Integral Non-Linearity (INL) is calculated by integrating the DNL, and the 12-bit raw outputs are corrected according to the INL; 3) After the INL correction, the last two bits are simply discarded to remove the missing codes, which are more uniformly distributed and a 10-bit accuracy is obtained. Figure~\ref{ADC_response}(b) shows the ADU value as function of voltage difference between the two differential input signals after calibration. The missing codes have been recovered by the simple calibration method and a better linear response achieved.

For the 10-bit output after calibration, the DNL and INL have been re-calculated using the afore-mentioned methods. It is found that, for the designs with different unit capacitance, the DNL for most of the codes are within $+/-$ 0.5 LSB, and the INL within $+/-$ 1 LSB; however, the DNL and INL for the codes at the two ends of the entire range are beyond their nominal values, which reduces the Effective Number Of Bits (ENOB) slightly < 10 (ENOB = 9.8 in ADC-0.2 prototype). This issue, which is not discussed in detail due to the limit of the context, has been found to be related to the settling of signals at the inputs of the comparator. This can be improved by increasing the "waiting time" before comparisons are made. Thus an adjustable delay for the bigger capacitors in the DAC array has been proposed and implemented in the ADC-0.3 prototype, with which a better DNL and INL over the entire range, as well as an ENOB of > 10 are expected.

%\begin{figure}
%\small
%\centering
%\caption{The DNL, INL and rampling curve of the ADC-0.2 prototype at 10-bit accuracy.}
%\includegraphics[width=71mm]{pic_dnl_and_inl.eps}
%\includegraphics[width=75mm]{pic_ramping_curve_10bit.eps}
%\label{ADC_calibration}
%\end{figure}

\section{Summary and discussion}

Gotthard-II is a silicon microstrip detector currently under development for the XFEL.EU. The analogue front-end of the Gotthard-1.7 prototype has been designed and tested. With a high DC gain design in the pre-amplifier, low capacitive coupling and negligible cross-talk have been achieved. The other characteristics, for example single-photon sensitivity and dynamic range, have been validated according to the specifications. Even with an acceptable noise for 12.4 keV X-ray photons, a new prototype has been designed aiming for noise optimization for the detection of X-ray photons with lower energies. Two versions of ADC prototypes have been designed with different architectures of their DAC arrays and investigated in this work: the ADC-0.1 prototype with a conventional DAC array only achieves a speed $\sim$ 10 MS/s; the ADC-0.2 prototype with a split-capacitor array achieves $\sim$ 25 MS/s but with missing codes which can be explained by the capacitance mismatch. After calibration to 10-bit using a simple method, the missing codes can be removed and a better linear response obtained. In addition, the ADC-0.2 prototype shows a very good noise performance. Based on the investigation, a new ADC prototype has been designed with adjustable side capacitance in order to solve the problem of the capacitance mismatch and tunable delay for a better signal settling.

%\acknowledgments

%This is the most common positions for acknowledgments. A macro is available to maintain the same layout and spelling of the heading.

%\paragraph{Note added.} This is also a good position for notes added after the paper has been written.


\end{document}